\documentstyle[epsf,12pt]{article}
\begin{document}

\catcode`@=11
\long\def\@caption#1[#2]#3{\par\addcontentsline{\csname
  ext@#1\endcsname}{#1}{\protect\numberline{\csname
  the#1\endcsname}{\ignorespaces #2}}\begingroup
    \small
    \@parboxrestore
    \@makecaption{\csname fnum@#1\endcsname}{\ignorespaces #3}\par
  \endgroup}
\catcode`@=12
\newcommand{\newc}{\newcommand}
\newc{\gsim}{\lower.7ex\hbox{$\;\stackrel{\textstyle>}{\sim}\;$}}
\newc{\lsim}{\lower.7ex\hbox{$\;\stackrel{\textstyle<}{\sim}\;$}}
\newc{\gev}{\,{\rm GeV}}
\newc{\mev}{\,{\rm MeV}}
\newc{\ev}{\,{\rm eV}}
\newc{\kev}{\,{\rm keV}}
\newc{\tev}{\,{\rm TeV}}
\newc{\mz}{m_Z}
\newc{\mpl}{M_{Pl}}
\newc{\chifc}{\chi_{{}_{\!F\!C}}}
\newc\order{{\cal O}}
\newc\CO{\order}
\newc\CL{{\cal L}}
\newc\CY{{\cal Y}}
\newc\CH{{\cal H}}
\newc\CM{{\cal M}}
\newc\CF{{\cal F}}
\newc\CD{{\cal D}}
\newc\CN{{\cal N}}
\newc{\eps}{\epsilon}
\newc{\re}{\mbox{Re}\,}
\newc{\im}{\mbox{Im}\,}
\newc{\invpb}{\,\mbox{pb}^{-1}}
\newc{\invfb}{\,\mbox{fb}^{-1}}
\newc{\yddiag}{{\bf D}}
\newc{\yddiagd}{{\bf D^\dagger}}
\newc{\yudiag}{{\bf U}}
\newc{\yudiagd}{{\bf U^\dagger}}
\newc{\yd}{{\bf Y_D}}
\newc{\ydd}{{\bf Y_D^\dagger}}
\newc{\yu}{{\bf Y_U}}
\newc{\yud}{{\bf Y_U^\dagger}}
\newc{\ckm}{{\bf V}}
\newc{\ckmd}{{\bf V^\dagger}}
\newc{\ckmz}{{\bf V^0}}
\newc{\ckmzd}{{\bf V^{0\dagger}}}
\newc{\X}{{\bf X}}
\newc{\bbbar}{B^0-\bar B^0}
\def\bra#1{\left\langle #1 \right|}
\def\ket#1{\left| #1 \right\rangle}
\newc{\sgn}{\mbox{sgn}\,}
\newc{\m}{{\bf m}}
\newc{\msusy}{M_{\rm SUSY}}
\newc{\munif}{M_{\rm unif}}
%
%
\def\NPB#1#2#3{Nucl. Phys. {\bf B#1} (19#2) #3}
\def\PLB#1#2#3{Phys. Lett. {\bf B#1} (19#2) #3}
\def\PLBold#1#2#3{Phys. Lett. {\bf#1B} (19#2) #3}
\def\PRD#1#2#3{Phys. Rev. {\bf D#1} (19#2) #3}
\def\PRL#1#2#3{Phys. Rev. Lett. {\bf#1} (19#2) #3}
\def\PRT#1#2#3{Phys. Rep. {\bf#1} (19#2) #3}
\def\ARAA#1#2#3{Ann. Rev. Astron. Astrophys. {\bf#1} (19#2) #3}
\def\ARNP#1#2#3{Ann. Rev. Nucl. Part. Sci. {\bf#1} (19#2) #3}
\def\MPL#1#2#3{Mod. Phys. Lett. {\bf #1} (19#2) #3}
\def\ZPC#1#2#3{Zeit. f\"ur Physik {\bf C#1} (19#2) #3}
\def\APJ#1#2#3{Ap. J. {\bf #1} (19#2) #3}
\def\AP#1#2#3{{Ann. Phys. } {\bf #1} (19#2) #3}
\def\RMP#1#2#3{{Rev. Mod. Phys. } {\bf #1} (19#2) #3}
\def\CMP#1#2#3{{Comm. Math. Phys. } {\bf #1} (19#2) #3}
\relax
%
%
%
\def\beq{\begin{equation}}
\def\eeq{\end{equation}}
\def\bea{\begin{eqnarray}}
\def\eea{\end{eqnarray}}
%
%
%
\newc{\ie}{{\it i.e.}}          \newc{\etal}{{\it et al.}}
\newc{\eg}{{\it e.g.}}          \newc{\etc}{{\it etc.}}
\newc{\cf}{{\it c.f.}}
\def\smuon{{\tilde\mu}}
\def\neut{{\tilde N}}
\def\char{{\tilde C}}
\def\bino{{\tilde B}}
\def\wino{{\tilde W}}
\def\higgsino{{\tilde H}}
\def\sneut{{\tilde\nu}}
\def\stau{{\tilde\tau}}
%
%
%
%
\def\slash#1{\rlap{$#1$}/} 
\def\Dsl{\,\raise.15ex\hbox{/}\mkern-13.5mu D} 
\def\delsl{\raise.15ex\hbox{/}\kern-.57em\partial}
\def\Ksl{\hbox{/\kern-.6000em\rm K}}
\def\Asl{\hbox{/\kern-.6500em \rm A}}
\def\Qsl{\hbox{/\kern-.6000em\rm Q}}
\def\gradsl{\hbox{/\kern-.6500em$\nabla$}}
%
%
%
\def\bar#1{\overline{#1}}
\def\vev#1{\left\langle #1 \right\rangle}
%

\begin{titlepage}
\begin{flushright}
UND-HEP-01-K03\\
August 2001\\
\end{flushright}
\vskip 2cm
\begin{center}
{\large\bf Implications of the Muon Anomalous Magnetic Moment \\
for Supersymmetry}
\vskip 1cm
{\normalsize\bf
Mark Byrne, Christopher Kolda and Jason E.~Lennon} \\
\vskip 0.5cm
{\it Department of Physics, University of Notre Dame\\
Notre Dame, IN~~46556, USA\\[0.1truecm]
}

\end{center}
\vskip .5cm

\begin{abstract}
We re-examine the bounds on supersymmetric particle masses in light of
the E821 data on the muon anomalous magnetic moment,
$a_\mu$. We confirm, extend and supersede previous bounds. In
particular we find (at $1\sigma$) 
no lower limit on $\tan\beta$ or upper limit on the chargino mass
implied by the data at present, but at least 4 sparticles must be 
lighter than 700 to $820\gev$ and at least one sparticle must be lighter 
than 345 to $440\gev$. However, the E821 central value bounds
$\tan\beta>4.7$ and $m_{\char_1}<690\gev$. For $\tan\beta\lsim10$, the
data indicates a high probability for direct discovery of SUSY at Run
II or III of the Tevatron.
\end{abstract}

\end{titlepage}

\setcounter{footnote}{0}
\setcounter{page}{1}
\setcounter{section}{0}
\setcounter{subsection}{0}
\setcounter{subsubsection}{0}


Recently, the Brookhaven E821 experiment~\cite{e821} has reported
evidence for a deviation of the muon magnetic moment from the 
Standard Model expectation. Immediately following that announcement 
appeared a number of papers analyzing the reported excess in terms of
various forms of new physics, including supersymmetry
(SUSY)~\cite{czar,kane,feng,martinwells,others}.
The SUSY explanation is, for many of us, the
most exciting of the various proposals since it implies SUSY at a mass
scale not far above the weak scale. In particular, it implies a light
slepton and a light gaugino, though ``light'' can still be as heavy as
several hundred GeV.

Previous analyses have generally concentrated on bounding the masses
of the chargino and the sneutrino, or on the lightest of the sparticles.
In this paper, we will re-examine the SUSY calculation, both in very
general scenarios and in well-motivated, simplified SUSY models. We
will show that there exist mass bounds on more than two SUSY
particles, though which sparticles are bounded changes with
$\tan\beta$. In particular, we will show that throughout parameter
space, one may actually bound {\it at least 4}\/ sparticle masses if
we are to use SUSY to explain the E821 data. The existence of these
bounds will rely on very simple and clearly stated assumptions about
the SUSY particle spectrum; these assumptions will not include a
fine-tuning constraint.

\section{SUSY and $a_\mu$}

The actual measurement performed by the E821 collaboration is of the
muon's {\it anomalous}\/ magnetic moment, which is to say, the
coefficient $a_\mu$ of the non-renormalizable operator 
$$\frac{a_\mu}{2m_\mu}\bar\psi\, \sigma^{\alpha\beta}\psi
F_{\alpha\beta}.$$
Within the Standard Model, $a_\mu$ is predicted to be~\cite{davier}
$$ a_\mu^{\rm SM} = 11\,659\,159.6(6.7)\times 10^{-10}.$$
It has been claimed that the quoted uncertainty is actually too 
small~\cite{yndurain};
the dominant uncertainty is the hadronic contribution to the photon
polarization diagrams which is extracted from experimental
measurements of $R(e^+e^-\to{\rm hadrons})$ in the vicinity of the
low-energy meson resonances. However, strong arguments have been made
to reinforce the quoted values~\cite{marciano}, 
and we will accept them here as given.

The measurement made by E821 is~\cite{e821}:
$$a_\mu^{\rm E821} = 11\,659\,202(14)(6)\times 10^{-10}$$
from which one deduces a discrepancy between the experiment and the
Standard Model of
$$\delta a_\mu = 43(16)\times 10^{-10},$$ 
that is, the measured value is {\it larger}\/ than the prediction by 
$2.7\sigma$.

The SUSY contributions to $a_\mu$ have been known since the early days
of SUSY and have become more complete with time\cite{calcs}. In this
paper we will follow the notation of Ref.~\cite{martinwells} which has the
advantage of using the standard conventions of Haber and
Kane~\cite{haberkane}; any convention which we do not define here can be
found in either of these two papers. 

Before we begin the discussion of our work, let us review briefly
a few of the analyses to date.
The first paper to use the new E821 data in the context of SUSY was that of
Czarnecki and Marciano~\cite{czar} who only attempted to approximately
bound the sparticle spectrum. More complete analyses followed quickly
thereafter by  Everett \etal~\cite{kane} and Feng and
Matchev~\cite{feng}. The former~\cite{kane} argued that only very
large values of $\tan\beta$ were generically consistent with the data;
given those, a $1.5\sigma$ upper bound of $450\gev$ could be
placed on some sparticle (gaugino or slepton) for $\tan\beta=35$,
and $900\gev$ for both a chargino and the muon sneutrino.
The latter analysis~\cite{feng} found a model-independent 
$1\sigma$ limit on the mass of the lightest ``observable'' sparticle of
$490\gev$ for $\tan\beta<50$. Martin and Wells did a more complete
analysis, but focussed on the lighter chargino and charged
smuon. Their model-independent analysis found no upper bound on the
chargino mass 
at $1\sigma$ but found $m_\smuon<500\gev$. With the added assumption
of gaugino mass unification (\ie, $M_2\simeq 2M_1$), the lighter
chargino was bounded by about $700\gev$ at $\tan\beta=30$.
The lack of a strong upper
bound on the chargino was due to the neutralino
contributions at small $\tan\beta$ (all the way down to $\tan\beta=3$)
which can explain the E821 data without any chargino piece at all. Finally,
there have been a number of other papers~\cite{others} 
which have studied the SUSY
contributions within specialized scenarios which we will not discuss here.

\subsection{The diagrams}

In the mass eigenbasis, there are only two one-loop SUSY 
diagrams which contribute to $a_\mu$, shown in Figure~1. The first has
an internal loop of smuons and neutralinos, the second a loop of
sneutrinos and charginos. But the charginos, neutralinos and even the
smuons are themselves admixtures of various interaction eigenstates
and we can better understand the physics involved by working in terms
of these interaction diagrams, of which there are many more than two.
We can easily separate the leading and subleading diagrams in the
interaction eigenbasis by a few simple observations.
\begin{figure}
\centering
\epsfxsize=4.25in
\hspace*{0in}
\epsffile{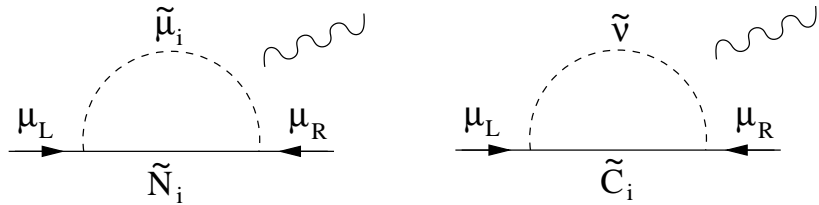}
\caption{Supersymmetric diagrams contributing to $a_\mu$ at one-loop.}
\label{simple}
\end{figure}

First, the magnetic moment operator is a helicity-flipping
interaction. Thus
any diagram which contributes to $a_\mu$ must involve a helicity flip
somewhere along the fermion current. This automatically divides the
diagrams into two classes: those with helicity flips on the external
legs and those with flips on an internal line. For those in the first
class, the amplitude must scale as $m_\mu$; for those in the second,
the amplitudes can scale instead by $m_{\rm SUSY}$, where $m_{\rm
SUSY}$ represents the mass of the internal SUSY fermion (a chargino or
neutralino). Since $m_{\rm SUSY}\gg m_\mu$, it is the latter class
that will typically dominate the SUSY contribution to
$a_\mu$. Therefore we will restrict further discussion to this latter
class of diagrams alone.

Secondly, the interaction of the neutralinos and charginos with the
(s)muons and sneutrinos occurs either through their higgsino or 
gaugino components. Thus each vertex implies a factor of either
$y_\mu$ (the muon Yukawa coupling) or $g$ (the weak and/or hypercharge
gauge coupling). Given two vertices, the diagrams therefore scale as
$y_\mu^2$, $gy_\mu$ or $g^2$.
In the Standard Model, $y_\mu$ is smaller than $g$ by
roughly $10^{-3}$. In the minmal SUSY standard model (MSSM) 
at low $\tan\beta$, this ratio is
essentially unchanged, but because $y_\mu$ scales as $1/\cos\beta$, at
large $\tan\beta$ ($\sim60$) the ratio can be reduced to roughly
$10^{-1}$. Thus we can safely drop the $y_\mu^2$ contributions from
our discussions, but at large $\tan\beta$ we must preserve the
$gy_\mu$ pieces as well as the $g^2$ pieces.\footnote{Pieces 
which are dropped from our discussion
are still retained in the full numerical calculation.}

The pieces that we will keep are therefore shown in Figures~2. In
Fig.~2(a)-(e) are shown the five neutralino contributions which scale 
as $g^2$
or $gy_\mu$; in Fig.~2(f) is the only chargino contribution, scaling as
$gy_\mu$. The contributions to $a_\mu$ from the $i$th
neutralino and the $m$th smuon due to each of these component
diagrams are found to be:
$$
\delta a_\mu = \frac{1}{48\pi^2}\frac{m_\mu m_{\neut_i}}{m_{\smuon_m}^2}
F_2^N(x_{im})
\times \left\lbrace \begin{array}{lc}
g_1^2 N_{i1}^2 X_{m1}X_{m2} & (\bino\bino)\\
g_1g_2 N_{i1}N_{i2} X_{m1}X_{m2} & (\wino\bino) \\
-\sqrt{2}g_1y_\mu N_{i1}N_{i3} X_{m2}^2 & (\higgsino\bino) \\
\frac{1}{\sqrt{2}}g_1y_\mu N_{i1}N_{i3} X_{m1}^2 & (\bino\higgsino)\\
\frac{1}{\sqrt{2}}g_2y_\mu N_{i2}N_{i3} X_{m1}^2 & (\wino\higgsino)
\end{array} \right.
$$
and for the $k$th chargino and the sneutrino:
$$\delta a_\mu = -\frac{1}{24\pi^2}\frac{m_\mu m_{\char_k}}{m_{\sneut}^2}
F_2^C(x_{k})\,g_2y_\mu U_{k2}V_{k1}.\quad\quad (\wino\higgsino)$$
The matrices $N$, $U$ and $V$ are defined in the appendix along with
the functions $F_2^{N,C}$. A careful comparison to the equations in
the appendix will reveal that we have dropped a number of complex
conjugations in the above expressions; it has been shown
previously~\cite{kane,martinwells} that the SUSY contributions to
$a_\mu$ are maximized for real entries in the mass matrices and so we
will not retain phases in our discussion.
\begin{figure}
\centering
\epsfxsize=4.25in
\hspace*{0in}
\epsffile{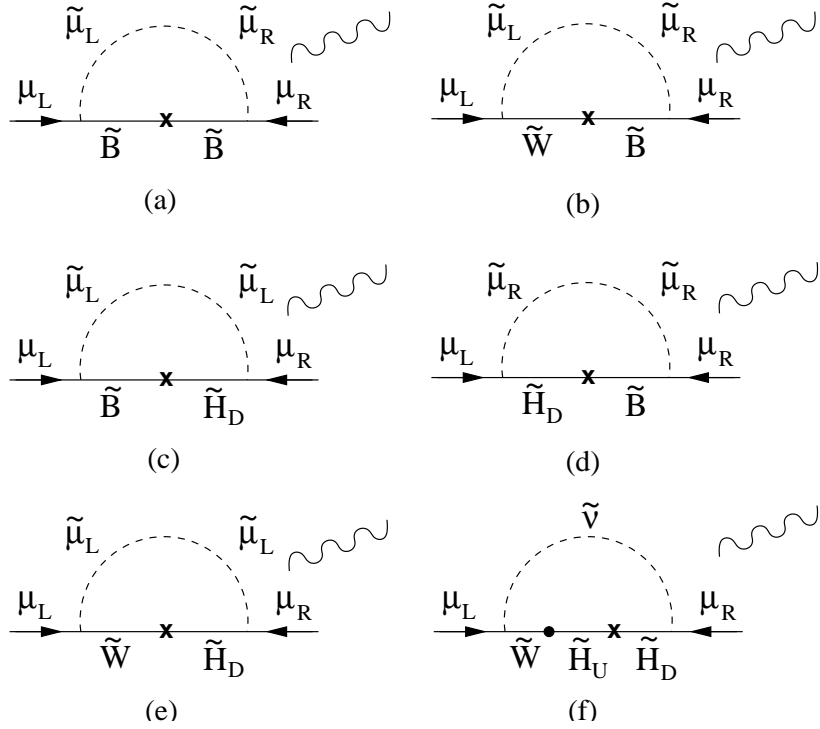}
\caption{Diagrams contributing to $a_\mu$ in the interaction eigenbasis.
}
\label{intdiags}
\end{figure}

In many of the previous analyses of the MSSM parameter space, it was
found that it is the chargino-sneutrino
diagram at large $\tan\beta$ 
that can most easily generate values of $\delta a_\mu$ large
enough to explain the observed discrepancy.
From this observation, one can obtain an upper mass bound on the
lightest chargino and the muon sneutrino. However, this behavior
is not completely generic. For example,
Martin and Wells have emphasized that the $\bino\bino$
neutralino contribution can {\it by itself}\/ be large enough to
generate the observed excess in $a_\mu$, and since it has no
intrinsic $\tan\beta$ dependence, they could explain the data with
$\tan\beta$ as low as 3. We can reproduce their result in a simple way
because the $\bino\bino$ contribution has a calculable upper bound at which
the smuons mix at $45^\circ$, $m_{\smuon_1}\ll m_{\smuon_2}$, and
$m_{\neut_1}\ll m_{\neut_{2,3,4}}$ with $\neut_1=\bino$. Then
$$\left|\delta a_\mu\right|_{(\bino\bino)}
\leq \frac{g_1^2}{32\pi^2}\frac{m_\mu
m_{\neut_1}}{m_{\smuon_1}^2}\simeq
3800\times10^{-10}\times \left(\frac{m_{\neut_1}}{100\gev}\right)
\left(\frac{100\gev}{m_{\smuon_1}}\right)^2$$
where we have used the fact that $(X_{11}X_{12})\leq\frac12$ and
$F_2^N\leq 3$ and have included a 7\% two-loop suppression factor.
Though any real model will clearly suppress this
contribution somewhat, this is still $10^2$ times larger than needed
experimentally. 

This pure $\bino\bino$ scenario is actually an experimental 
worst-case, particularly for hadron colliders. The only
sparticles that are required to be light are a single neutralino
(which is probably $\bino$-like) and a single $\smuon$. The neutralino
is difficult to produce, and if stable, impossible to detect
directly. The neutralino could be indirectly observed in the decay of
the $\smuon$ as missing energy, but production of a $\smuon$ at a
hardon machine is highly suppressed. In the worst of all possible
worlds, E821 could be explained by only these two light sparticles,
with the rest of the SUSY spectrum hiding above a TeV. Further, even
the ``light'' sparticles can be too heavy to produce at a $500\gev$
linear collider. While this case
is in no way generic, it demonstrates that the E821 data by itself
does not provide any sort of no-lose theorem for Run~II of the
Tevatron or even for the LHC and NLC.

This raises an important experimental question: how many of the MSSM 
states must be ``light'' in order to explain the E821 data? In the
worst-case, it would appear to be only two.
Even in the more optimistic scenario in which the chargino diagram
dominates $\delta a_\mu$, the answer naively appears to be two: 
a single chargino and a single sneutrino. In this limit, 
$$\left|\delta a_\mu\right|_{(\char\sneut)}
\leq \frac{g_2 y_\mu}{24\pi^2}\frac{m_\mu m_{\char_1}}{m_\sneut^2}
\left|F^C_2\right|_{\rm max} \lsim 2600\times 10^{-10}\times
\left(\frac{m_{\char_1}}{100\gev}\right)
\left(\frac{100\gev}{m_{\sneut}}\right)^2
\left(\frac{\tan\beta}{30}\right)$$
where we have bounded $|F_2^C|$ by 10 by assuming $m_\sneut\lsim 1\tev$.
But this discussion is overly simplistic and we can do much
better, as we will see.

\subsection{Mass correlations}

There are a total of 9 separate sparticles which can enter the
loops in Fig.~\ref{simple}: 1 sneutrino, 2 smuons, 2 charginos and 4
neutralinos. The mass spectrum of these 9 sparticles is determined
entirely by 7 parameters in the MSSM: 2 soft slepton masses($m_L$,
$m_R$), 2 gaugino masses ($M_1$, $M_2$), the $\mu$-term, a soft
trilinear slepton coupling ($A_\smuon$) and finally $\tan\beta$. Of
these, $A_\smuon$ plays almost no role at all and so we leave it out of
our discussions (see the appendix). And in some well-motivated
SUSY-breaking scenarios, $M_1$ and $M_2$ are also correlated. Thus
there are either 5 or 6 parameters responsible for setting 9 sparticle
masses. There are clearly non-trivial correlations among the
masses which can be exploited in setting mass limits on the sparticles.

First, there are well-known correlations between the
chargino and neutralino masses; for example, 
a light charged $\char_i\sim\wino$
implies a light neutral $\neut_j\sim\wino$ and vice-versa. 

There are also
correlations in mixed systems (\ie, the neutralinos, charginos and
smuons) between the masses of the eigenstates and the size of their
mixings. Consider the case of the smuons in particular; their mass
matrix is given in the appendix. On diagonalizing, the left-right
smuon mixing angle is given simply by:
$$\tan 2\theta_\smuon\simeq\frac{2 m_\mu\mu\tan\beta}{M_L^2-M_R^2}.$$
The chargino contribution is maximized for large smuon mixing and
large mixing occurs when the numerator is of order or greater 
than the denominator; since the former is suppressed by $m_\mu$, one
must compensate by having either a very large $\mu$-term in the numerator or 
nearly equal $M_L$ and $M_R$ in the denominator, both of which have
profound impacts on the spectrum.


There is one more correlation/constraint that we feel is natural to
impose on the MSSM spectrum: slepton mass universality. It is
well-known that the most general version of the MSSM produces huge
flavor-changing neutral currents (FCNCs) unless some external order is
placed on the MSSM spectrum. By far, the simplest such order is the demand
that sparticles with the same gauge quantum numbers be degenerate. Thus
we expect $m_{\stau_L}=m_{\smuon_L}\equiv m_L$ and
$m_{\stau_R}=m_{\smuon_R}\equiv m_R$. 
Then the mass matrix for the stau sector
is identical to that of the smuons with the replacement $m_\mu\to
m_\tau$ in the off-diagonal elements. This enhancement of the mixing
in the stau sector by $m_\tau/m_\mu\simeq 17$ implies that
$m_{\stau_1} < m_{\smuon_1}$. In particular, if 
$$M_L^2 M_R^2<m^2_\tau\,\mu^2\tan^2\beta$$ 
then $m^2_{\stau_1}<0$ and QED will be broken by a stau vev. Given
slepton universality, this imposes a constraint on the smuon mass
matrix:
$$M_L^2 M_R^2 > \left(\frac{m_\tau}{m_\mu}\right)^2 m_\mu^2\,\mu^2
\tan^2\beta$$ 
or on the smuon mixing angle:
$$\tan2\theta_\smuon < \left(\frac{m_\mu}{m_\tau}\right)
\frac{2M_LM_R}{M_L^2-M_R^2}$$
where $M_{L,R}$ are the positive roots of $M^2_{L,R}$. While not
eliminating the possibility of $\theta_\smuon\simeq 45^\circ$, 
this formula shows
that a fine-tuning of at least 1 part in 17 is needed to obtain
$\CO(1)$ mixing. We will not apply any kind of fine-tuning
criterion to our analysis, yet we will find that this slepton mass
universality constraint sharply reduces the upper bounds on slepton
masses which we are able to find in our study of points in MSSM
parameter space.

(As an aside, if one assumes slepton mass universality at some
SUSY-breaking messenger scale above the weak scale, 
Yukawa-induced corrections will
break universality by driving the stau masses down. This effect would
further tighten our bounds on smuon masses and mixings.)

The above discussion has an especially large impact on the worst-case
scenario in which the $\bino\bino$ contibutions dominates $\delta a_\mu$.
For generic points in MSSM parameter space,
one expects that $\tan 2\theta\lsim1/17$ which reduces the
size of the $\bino\bino$ contribution by a factor of 17. As a
byproduct, the masses required for explaining the E821 anomaly are
pushed back towards 
the range that can be studied by a $500\gev$ linear collider.

\section{Numerical results}

Now that we have established the basic principle of our analysis, let
us carry it out in detail. We will concentrate on three basic
cases. The first case is the one most often considered in the literature:
gaugino mass unification. Here one assumes that the weak-scale gaugino
mass parameters ($M_1$ and $M_2$) are equal at the same scale at which
the gauge couplings unify. This implies that at the weak scale
$M_1=(5/3)(\alpha_1/\alpha_2)M_2$. The second case we consider is
identical to the first with the added requirement that the lightest
SUSY sparticle (LSP) be a neutralino. This requirement is motivated by
the desire to explain astrophysical dark matter by a stable
LSP. Finally we will also consider the most general case in which all
relevant SUSY parameters are left free independent of each other; we
will refer to this as the ``general MSSM'' case.

The basic methodology is simple: we put down a
logarithmic grid on the space of MSSM parameters ($M_1$, $M_2$, $m_L$,
$m_R$ and $\mu$) for several choices of $\tan\beta$. The grid extends
from $10\gev$ for $M_1$, $M_2$ and $\mu$, and from $50\gev$ for
$m_{L,R}$, up to $2\tev$ for all mass parameters. For the case in which
gaugino unification is imposed $M_2$ is no longer a free parameter and
our grid contains $10^8$ points. For the general MSSM case
our grid contains $3\times10^9$ points. Only $\mu>0$ is considered since
that maximizes the value of $\delta a_\mu$. Finally, for our limits on
$\tan\beta$ we used an adaptive mesh routine which did a better job of
maximizing $\delta a_\mu$ over the space of MSSM inputs. By running
with grids  of varying resolutions and offsets we estimate the error
on our mass bounds to be less than $\pm 5\%$.

\subsection{Bounds on the lightest sparticles}

Perhaps the most important information that can be garnered from the
E821 data is an upper bound on the scale of sparticle masses. In
particular, one can place upper bounds on the masses of the lightest
sparticle(s) as a function of $\delta a_\mu$. Previous analyses have
often followed this approach, deriving upper bounds on the lightest
sparticle from among the gauginos and sleptons, or even more
specifically, from among the charginos and smuons. We too derive
bounds on the lightest sparticles, but as we have argued in the
previous section, we can also derive bounds on the second, third, and
even fourth lightest sparticles through the mass correlations.

These bounds on additional light sparticles provide an important
lesson. Without them there remains the very real possibility that the
E821 data is explained by a pair of light sparticles and that the
remaining SUSY spectrum is out of reach experimentally. But our
additional bounds
will give us some indication not only of where we can find SUSY, but
also of how much information we might be able to extract about the
fundamental parameters of SUSY --- the more sparticles we detect and
measure, the more information we will have for disentangling the
soft-breaking sector of the MSSM.

\begin{figure}
\centering
\parbox{3.05truein}{
\epsfxsize=3in
\hspace*{0in}
\epsffile{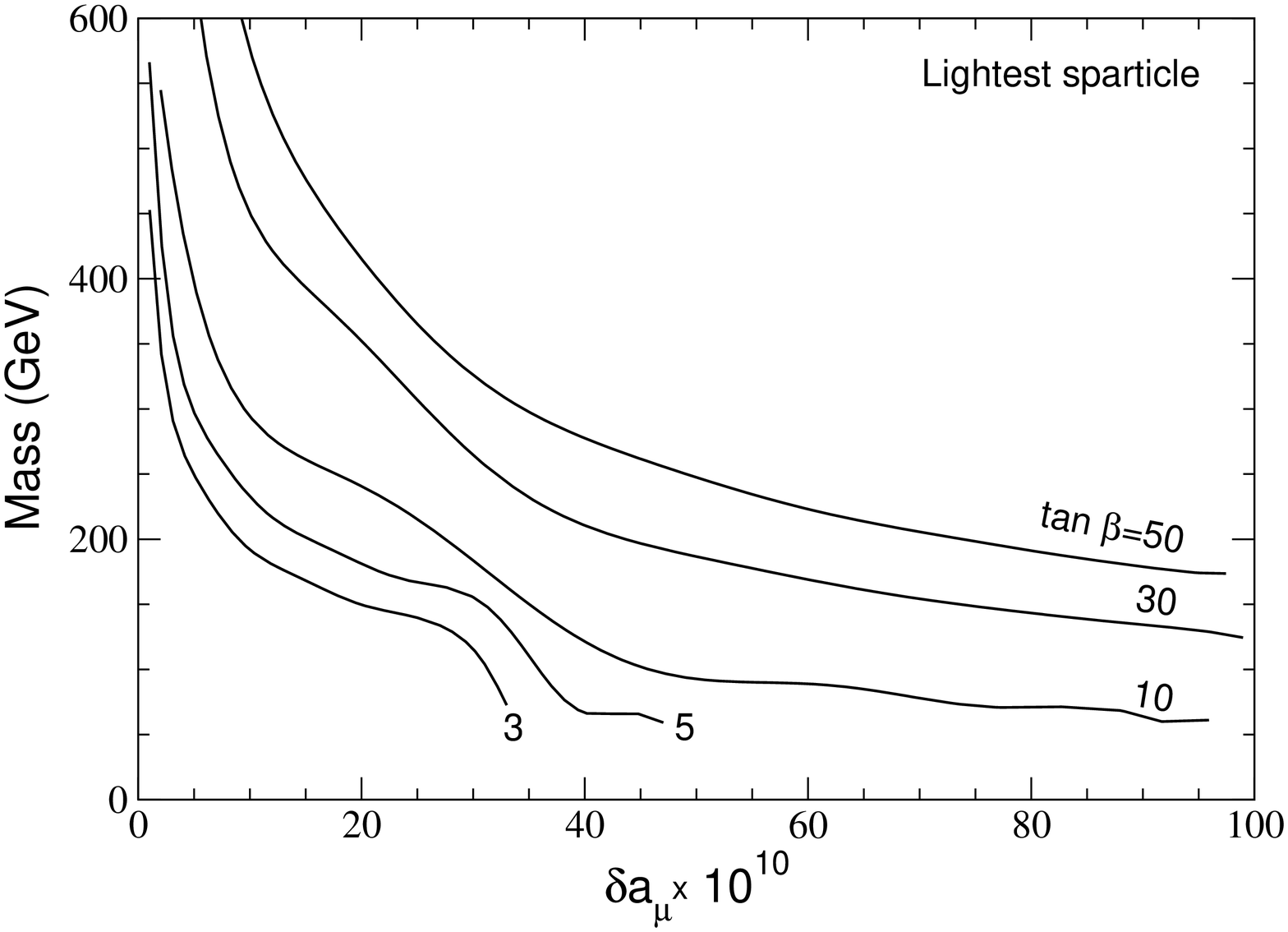}
}~\parbox{3.05truein}{
\epsfxsize=3.0in
\hspace*{0in}
\epsffile{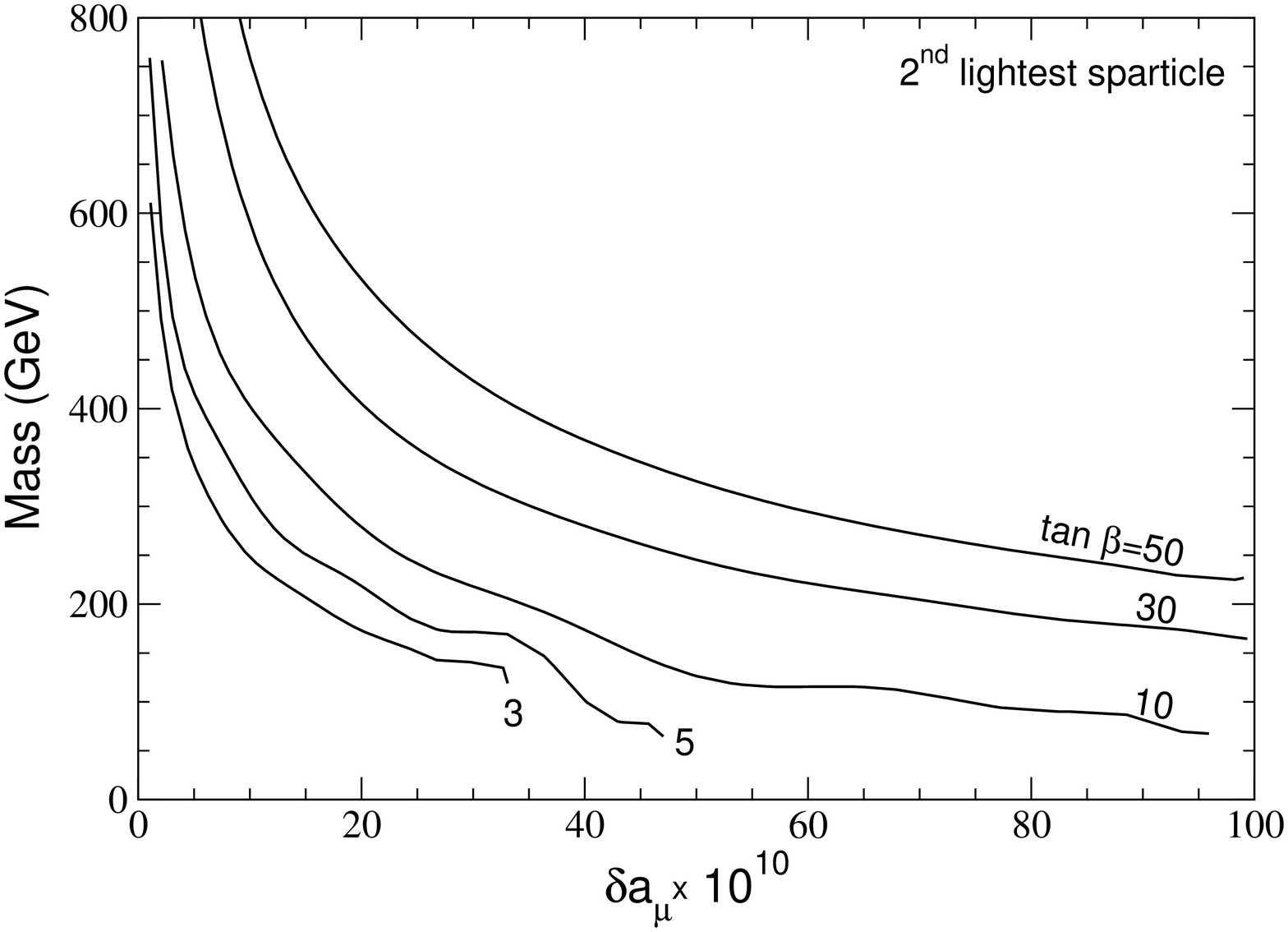}
}
\parbox{5truein}{~~~\\~~~~\\}
\parbox{3.05truein}{
\epsfxsize=3.0in
\hspace*{0in}
\epsffile{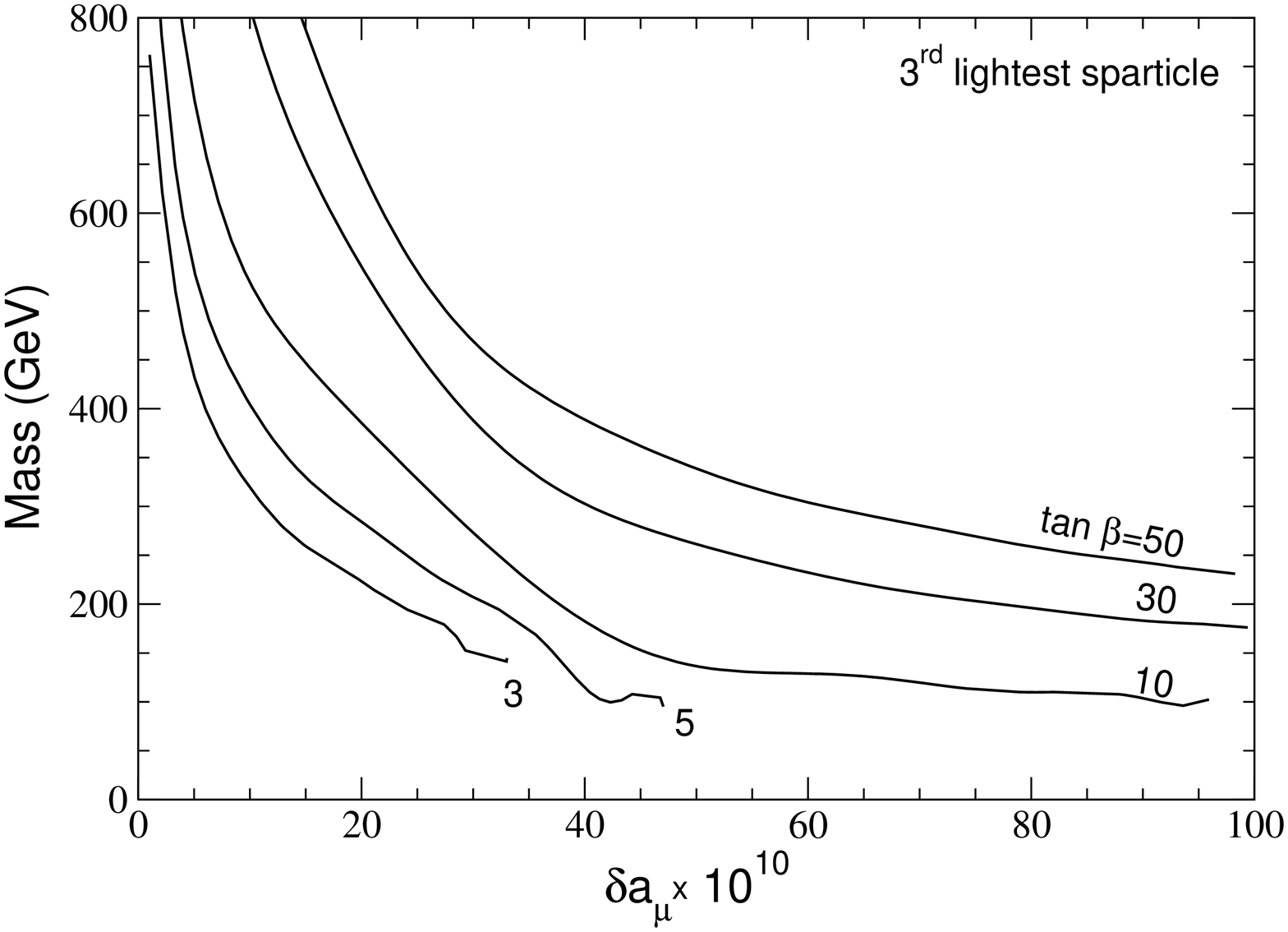}
}~\parbox{3.05truein}{
\epsfxsize=3.0in
\hspace*{0in}
\epsffile{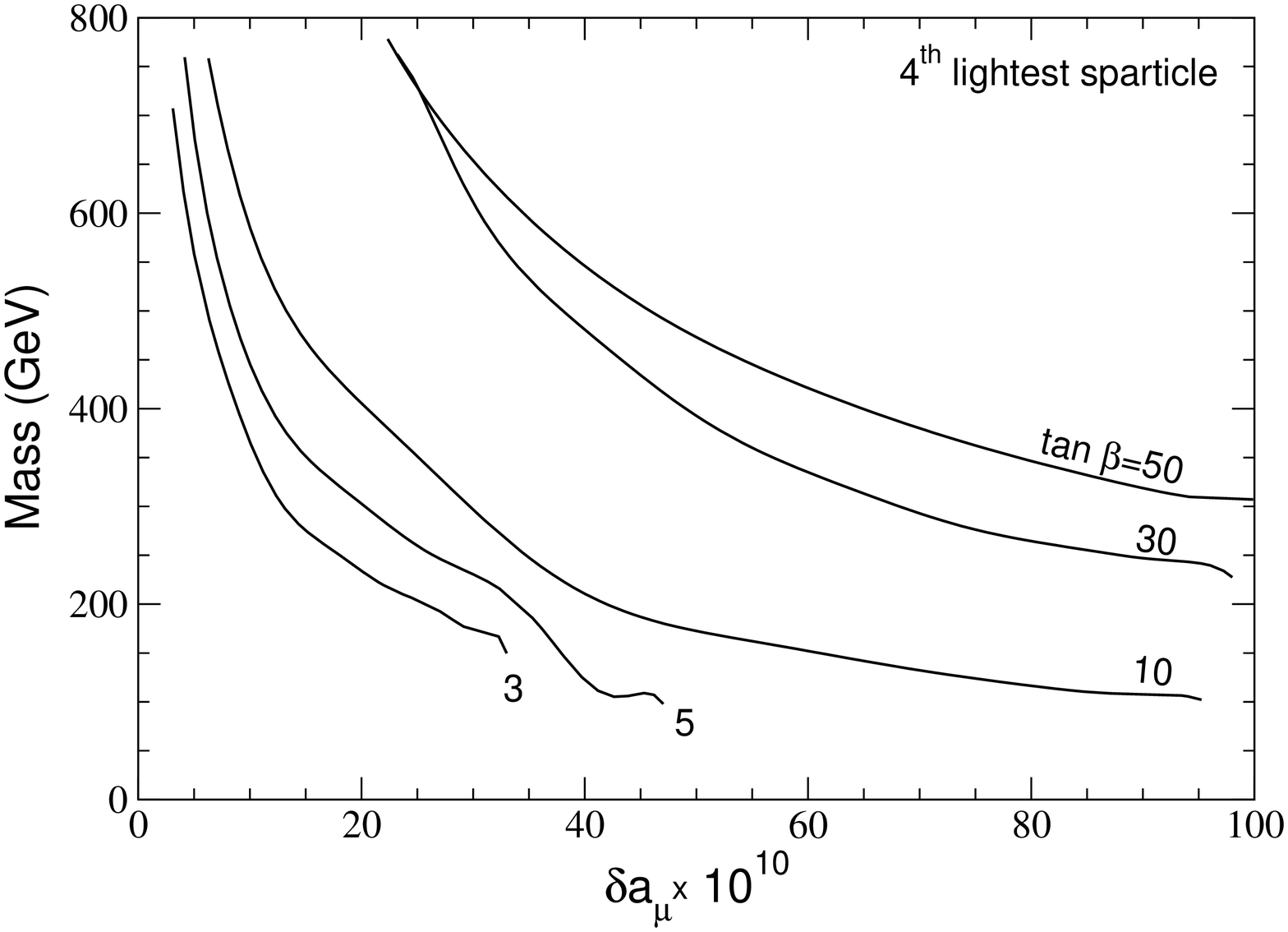}
}
\caption{Bounds on the masses of the four lightest sparticles as a
function of $\delta a_\mu$ for $\tan\beta=3$, 5, 10, 30 and 50.
These figures assume gaugino unification only.
}
\label{massfig}
\end{figure}

In Fig.~\ref{massfig} we have shown the upper mass bounds for the
lightest four sparticles assuming gaugino mass unification. 
These bounds are not bounds on individual species of sparticles (which
will come in the next section and always be larger than these bounds)
but simply bounds on whatever sparticle happens to be lightest.
The important points to note are: {\it (i)}\/ the maximum values of
the mass correspond to the largest value of $\tan\beta$, which is to
be expected given dominance of the chargino diagram at large
$\tan\beta$; {\it (ii)}\/ the $1\sigma$ limit (central value) 
of the E821 data requires
{\sl at least}\/ 4 sparticles to lie below roughly $700\,(500)\gev$; 
and {\it (iii)}\/ for low values of $\tan\beta$ a maximum
value of $\delta a_\mu$ is reached (we will return to this later).

\begin{figure}
\centering
\epsfxsize=4.25in
\hspace*{0in}
\epsffile{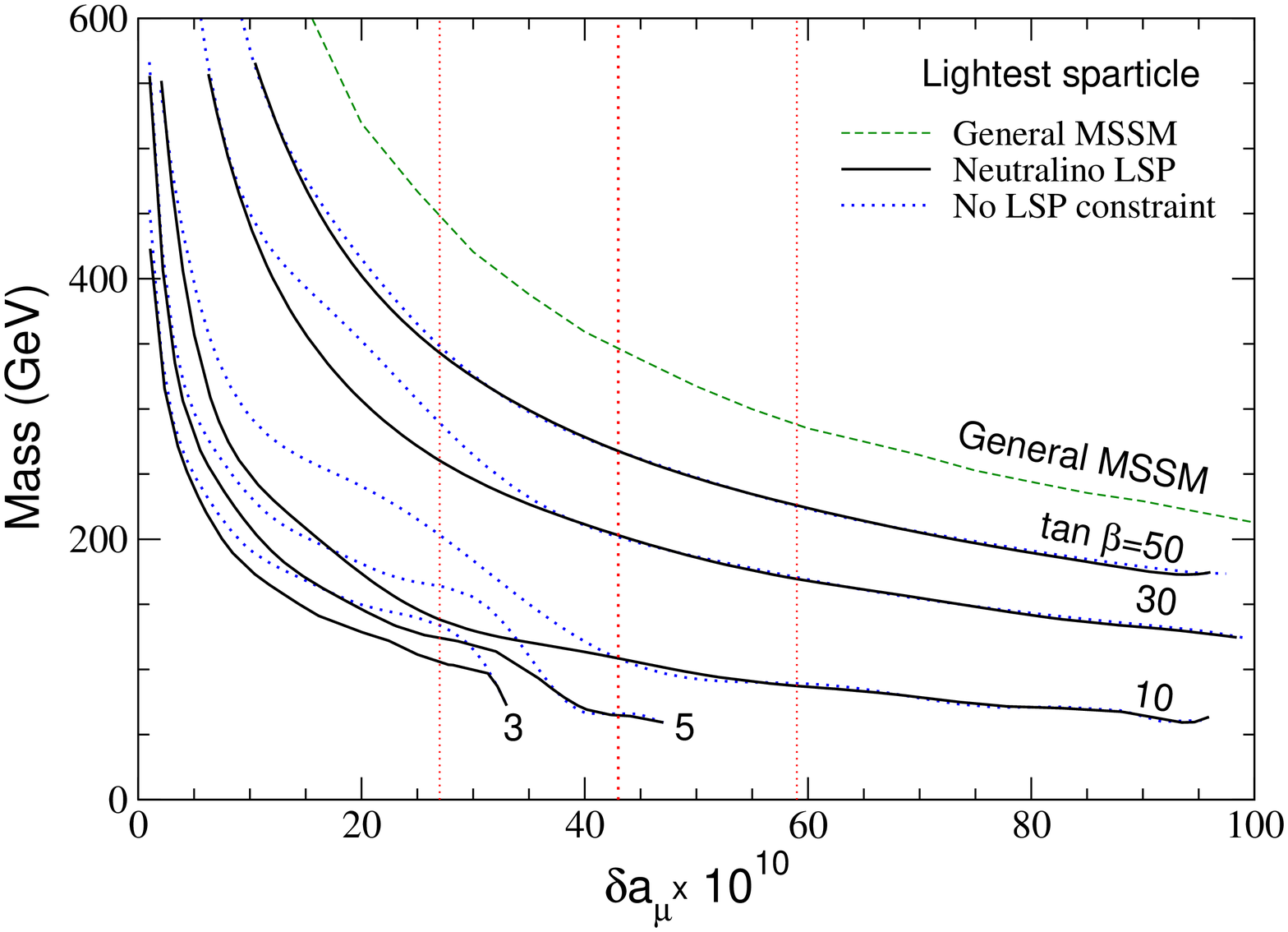}
\caption{Bound on the mass of the LSP as a
function of $\delta a_\mu$ for $\tan\beta=3$, 5, 10, 30 and 50. The 
dotted lines assume gaugino unification only while the solid lines
require additionally that the LSP is a neutralino. The dashed line is
the bound in the general MSSM, calculated at $\tan\beta=50$.
}
\label{LSPfig}
\end{figure}

The same plots could be produced with the additional assumption that
the LSP be a neutralino, but we will only show the case for the LSP
bound, in Fig.~\ref{LSPfig}. In this figure, the solid lines
correspond to a neutralino LSP, while the dotted lines are for the
more general case discussed above (\ie, they match the lines in
Fig.~\ref{massfig}(a)). Notice that for $\delta a_\mu\gsim 40\times
10^{-10}$ there is little difference between the cases with and
without a neutralino LSP. Furthermore, at the extreme upper and lower
values of $\tan\beta$ there is little difference. It is only for the
intermediate values of $\tan\beta$ that the mass bound shifts
appreciably; for $\tan\beta=10$ it comes down by as much as $50\gev$
compared to the more general case.

\begin{figure}
\centering
\epsfxsize=4.25in
\hspace*{0in}
\epsffile{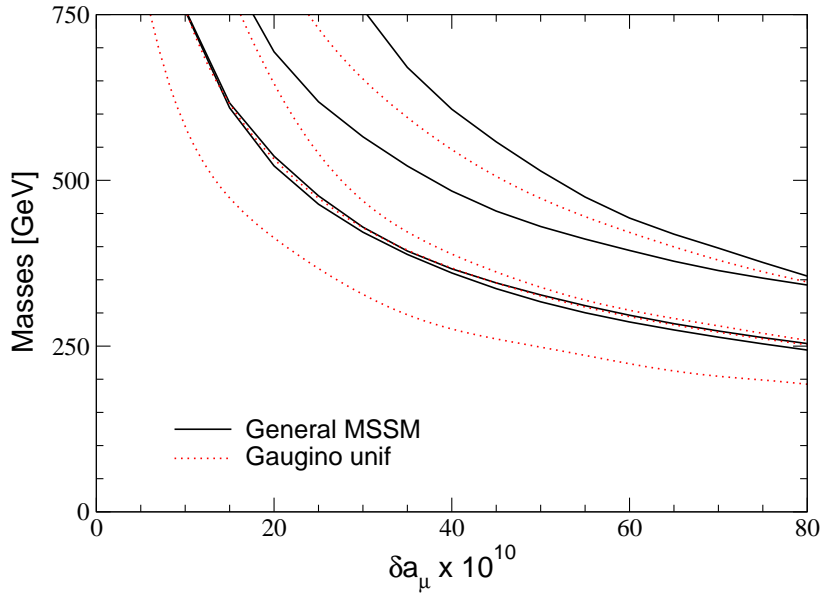}
\caption{Bounds on the masses of the four lightest sparticles as a 
function of $\delta a_\mu$ for $\tan\beta=50$. The 
dotted lines assume gaugino unification only while the solid lines
are for the general MSSM.
}
\label{nongutmaxes}
\end{figure}

Finally, we consider the most general MSSM case, \ie,  
without gaugino unification. Here the correlations are much less
pronounced, but interesting bounds still exist. For example the
central value of the E821 data still demands at least 3 sparticles
below $500\gev$ (rather than four for the previous cases).
In figure~\ref{nongutmaxes} we demonstrate this explicitly by
plotting the masses of the four lightest sparticles for $\tan\beta=50$
and a wide range of $\delta a_\mu$. We see that dropping the gaugino
unification requirement has one primary effect: the mass of the LSP is
significantly increased. This is because the LSP in the unified case
is usually a $\neut_1\sim\bino$ but isn't itself responsible for
generating $\delta a_\mu$. In the general case, the LSP must
participate in $\delta a_\mu$ (otherwise its mass could be arbitrarily
large) and so is roughly the mass of the {\it second}\/ lightest
sparticle in the unified case, whether that be a $\smuon$ or $\char$.
Otherwise the differences between the more general MSSM and the
gaugino unified MSSM are small. In particular we still find that at
least 4 sparticles must be light, though the $1\sigma$ bound of
$700\gev$ for the unified case extends now slightly to $820\gev$.

We have summarized all this data on the LSP in Table~\ref{masstable}
where we have shown the mass bounds (using both the $1\sigma$ limit
and the central value of the E821 data) on the LSP for various
$\tan\beta$ values and with our various assumptions. The last line in
the table represents an upper bound for any model with $\tan\beta\leq50$:
$m_{\rm LSP}<440\,(345)\gev$ for the E821
$1\sigma$ lower bound (central value) of $\delta a_\mu$.
But perhaps of equal importance are the bounds on the next 3 lightest
sparticles (the ``2LSP,'' ``3LSP'' and ``4LSP''): $m_{\rm 2LSP}<
460\,(355)\gev$, $m_{\rm 3LSP}< 600\,(465)\gev$ and $m_{\rm 4LSP}<
820\,(580)\gev$.
Thus there must be at least 4 sparticles below $820\gev$ even in the
most general MSSM scenario, and at least two sparticles accessible to
a $\sqrt{s}=1\tev$ linear collider.
\begin{table}
\centering
\begin{tabular}{r||c|c|c|}
\multicolumn{1}{c||}{Mass} & General & Gaugino & + Dark \\
\multicolumn{1}{c||}{Bound} & MSSM & Unification & Matter \\ \hline\hline
$\tan\beta=3$ & 140 {\sl (~)}& 135 {\sl (~)}& 105 {\sl (~)}\\
5 & 165 {\sl (75)}& 160 {\sl(65)}& 125 {\sl(65)}\\
10 & 215 {\sl(135)}& 210 {\sl(105)}& 140 {\sl(105)} \\
30 & 335 {\sl (255)}& 285 {\sl(205)}& 260 {\sl(205)}\\
50 &\fbox{{\bf 440} {\sl(345)}}& {\bf 345} {\sl(265)}& {\bf 345} {\sl(265)}\\
\cline{2-4}
\end{tabular}
\caption{Upper bounds on the mass (in GeV) of the
lightest sparticle for the
general MSSM, the MSSM assuming gaugino mass unification, and the MSSM
with gaugino mass unification plus a neutralino LSP. The entries
represent the bound for the $1\sigma$ limit {\sl (central value)} of the
E821 data. The boldfaced
$\tan\beta=50$ entries represent upper bounds over all $\tan\beta\leq
50.$ Missing entries occur where the central value of the data is
unobtainable; see Fig.~\protect\ref{massfig}.
}
\label{masstable}
\end{table} 

\subsection{Bounds on the sparticle species}

In the previous subsection, we derived bounds on the lightest
sparticles, independent of the identity of those sparticles. Another
important piece of information that can be provided by this analysis
is bounds on individual species of sparticles, for example, on the
charginos or on the smuons. These bounds will of mathematical
necessity be higher than those derived in the previous section, but
still provide important information about how and where to look for
SUSY. In particular, they can help us gauge the likelihood
of finding SUSY at Run~II of the Tevatron or at the LHC.

There is one complication in obtaining these bounds. At low
$\tan\beta$ the data is most easily explained by the neutralino
diagrams and as such there must be at least one light smuon and one
light neutralino. At larger $\tan\beta$ values of $\delta a_\mu$ as
large as the E821 central value generally require contributions from
the chargino diagrams, so there must be a light chargino and a light
sneutrino. However the correlations already discussed preserve the
bounds on the various species over the whole range of $\tan\beta$. A
bound on $m_\sneut$ implies a bound on $m_{\smuon_1}$, and a bound on
$m_{\char_1}$ implies a bound on at least one of the $m_{\neut_i}$,
and in certain cases (such as gaugino unification), 
the converses may be true as well.

\begin{figure}
\centering
\parbox{3truein}{
\epsfxsize=3in
\hspace*{0in}
\epsffile{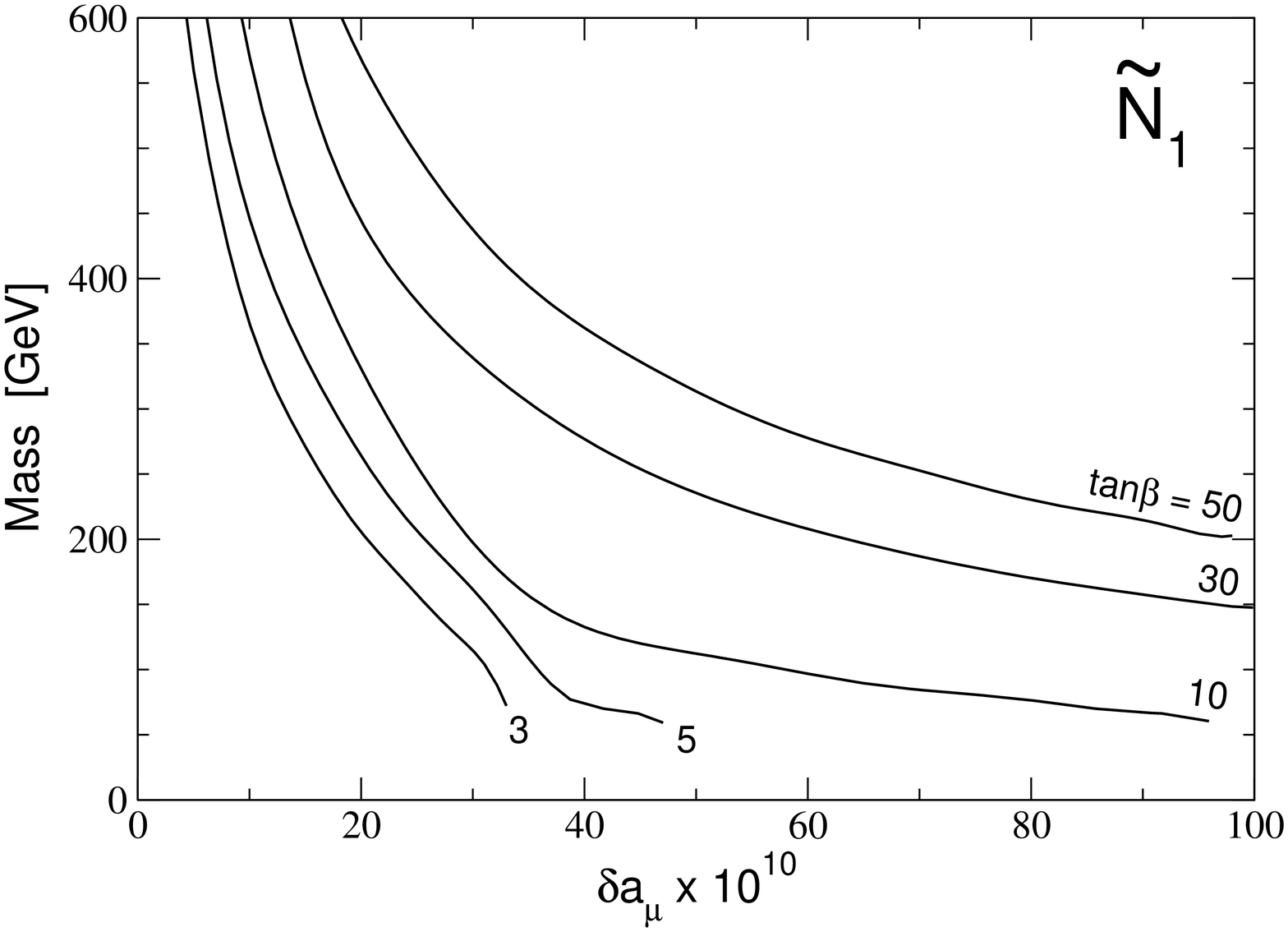}
}~~\parbox{3truein}{
\epsfxsize=3in
\hspace*{0in}
\epsffile{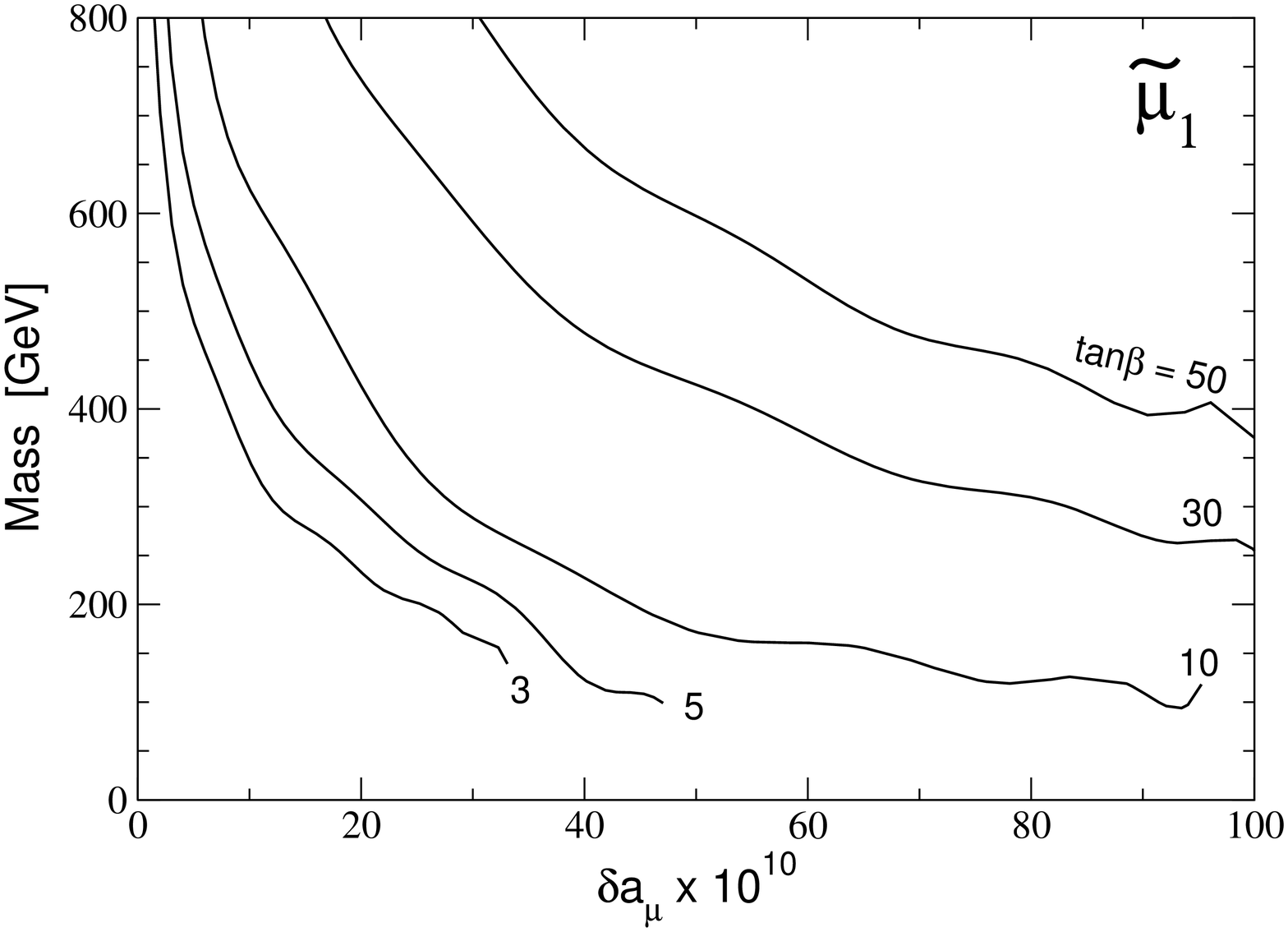}}
\caption{Bounds on the masses of $\neut_1$ and $\smuon_1$
as a function of $\delta a_\mu$ for various $\tan\beta$ with 
gaugino unification assumed.
}
\label{species}
\end{figure}

We have shown in Fig.~\ref{species} the mass bounds on $\smuon_1$ and
$\neut_1$ under the assumption of gaugino unification; a plot for 
$\char_1$/$\neut_2$ will appear later in our discussion of Tevatron
physics. Note that $\neut_1$ must be relatively light, even for large
$\tan\beta$, thanks to the gaugino unification condition, while
$\smuon_1$ can be heavier but must still lie below $820\gev$ at $1\sigma$.

\begin{figure}
\centering
\epsfxsize=4.5in
\hspace*{0in}
\epsffile{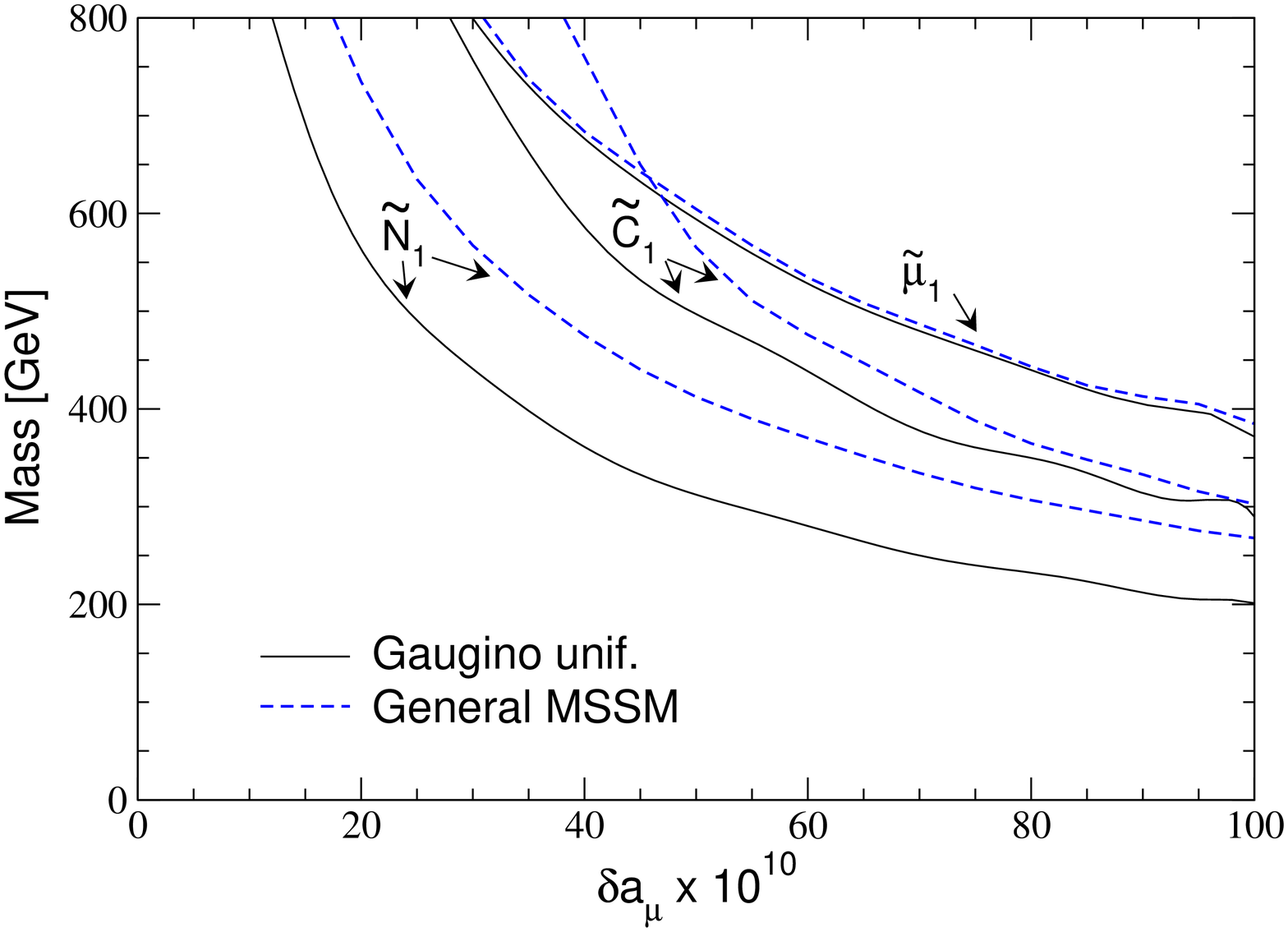}
\caption{Bounds on the masses of $\smuon_1$, $\char_1$,
and $\neut_1$
as a function of $\delta a_\mu$ for $\tan\beta=50$, with
gaugino unification (solid) and in the general MSSM (dashed).
The two lines for the $\smuon_1$ essentially overlap.
}
\label{nogut}
\end{figure}

Finally, we can consider the general MSSM without gaugino
unification. The results can best be summarized by showing the mass
bounds on the various sparticles at $\tan\beta=50$ in relations to
their bounds in the unified case. This is done in Fig.~\ref{nogut}.
We can see from the figure that the bound on the $\smuon_1$ is
essentially identical to that in the gaugino unification
picture. However the gaugino masses have shifted, and the reason is
no mystery. Once again, as discussed in the previous section, the
lightest neutralino is no longer a $\bino$-like spectator to the
magnetic moment, but is a $\wino$-like partner of a participating
$\wino$-like chargino.

We summarize our results for the various sparticle species in
Table~\ref{masseachtable}. There we have shown the upper bounds on
several sparticles in the general MSSM, the MSSM with gaugino
unification, and the previous case with the additional requirement of
a neutralino LSP (``dark matter''). The bounds represents those
obtained using the $1\sigma$ limit (central value) of the E821 data. 
\begin{table}
\centering
\begin{tabular}{c||c|c|c|}
\multicolumn{1}{c||}{Mass} & General & Gaugino & + Dark \\
\multicolumn{1}{c||}{Bound} & MSSM & Unification & Matter \\ \hline\hline
$\neut_1$ & 610 {\sl (455)}& 470 {\sl (350)}& 340 {\sl (265)}\\
$\neut_2$ & none {\sl (none)}& 830 {\sl(565)}& 575 {\sl(440)}\\
$\char_1$ &  none {\sl(690)}& 830 {\sl(565)}& 575 {\sl(440)} \\
$\smuon_1$ & 870{\sl (680)}& 825 {\sl(665)}& 825 {\sl(665)}\\
$\sneut$ & 865 {\sl(675)}& 820 {\sl(660)}& 820 {\sl(660)}\\
\cline{2-4}
\end{tabular}
\caption{Upper bounds on the mass (in GeV) of various
sparticles for the
general MSSM, the MSSM assuming gaugino mass unification, and the MSSM
with gaugino mass unification plus a neutralino LSP. The entries
represent the bound for the $1\sigma$ limit {\sl (central value)} of the
E821 data. These bounds are for all $\tan\beta\leq50$.
}
\label{masseachtable}
\end{table} 

\subsection{Bounds on $\tan\beta$}

\begin{figure}
\centering
\epsfxsize=4.25in
\hspace*{0in}
\epsffile{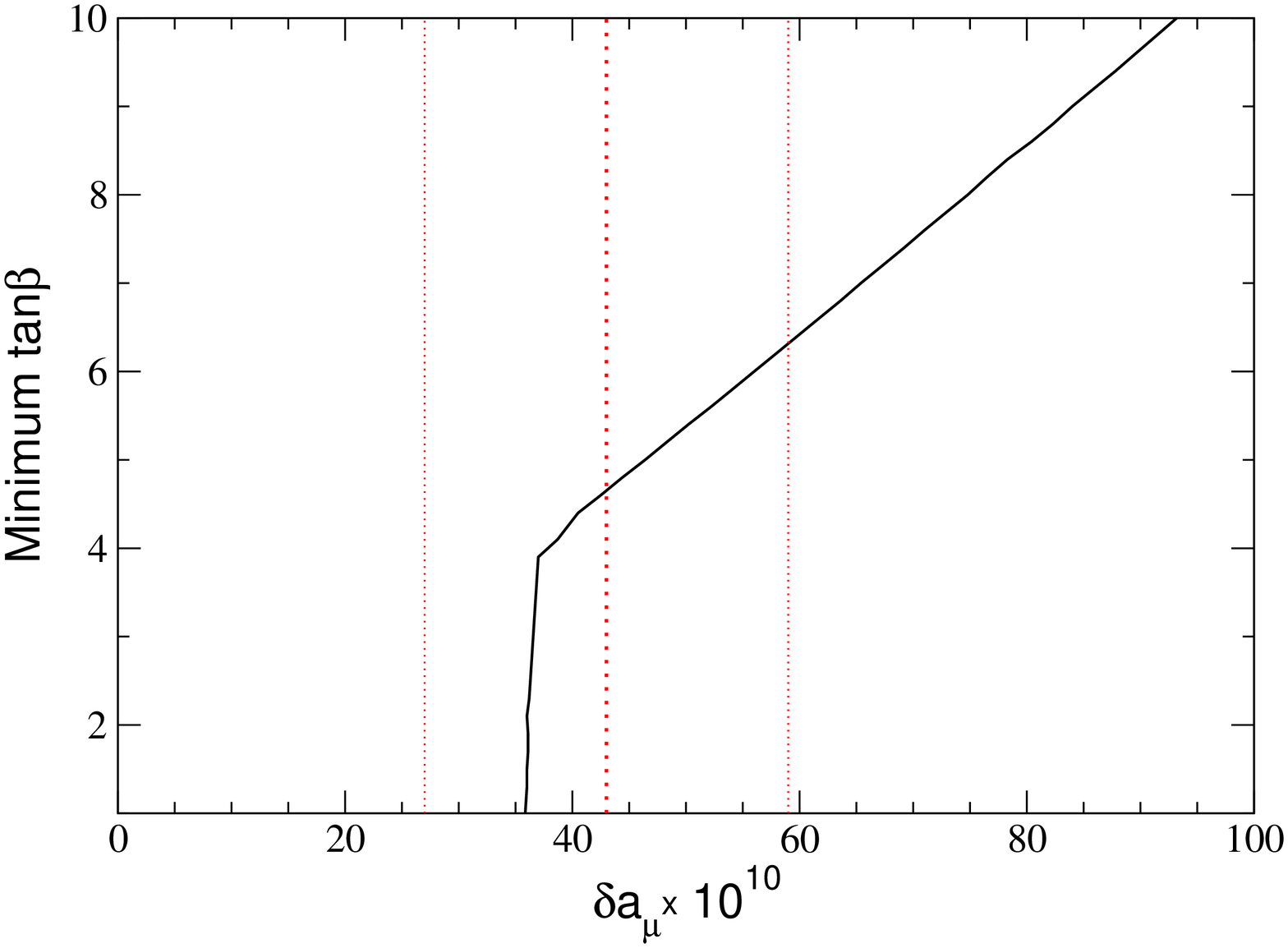}
\caption{Bounds on $\tan\beta$ as a function of $\delta a_\mu$ for the
general MSSM. The
dotted lines represent the E821 central value and $\pm1\sigma$ bounds
on $\delta a_\mu$.
}
\label{tanb}
\end{figure}

The final bound we will derive using the E821 data is on
$\tan\beta$. There has been some discussion in the literature about
which values of $\tan\beta$ are capable of explaining the data. And in
fact, we concur that at lower $\tan\beta$, there is a real suppression
in the maximum size of $\delta a_\mu$. So in Figure~\ref{tanb} we have
shown the maximum attainable value of $\delta a_\mu$ as a function of
$\tan\beta$ with and without the added assumption of gaugino mass
unification. 

The limit in Fig.~\ref{tanb} clearly divides into two regions. At
$\delta a_\mu>36\times 10^{-10}$ the chargino contribution
dominates and thus $\delta a_\mu\propto y_\mu$, scaling
linearly with $\tan\beta$. At lower $\delta a_\mu$, however, both
neutralino and chargino contributions can be important so it becomes
possible to generate $\delta a_\mu$ with much smaller values of
$\tan\beta$ than would be possible from the charginos alone. The
central value of the E821 data implies $\tan\beta>4.7$; but
already at $1\sigma$ all values of $\tan\beta$ down to 1 are allowed.
This result contradicts most statements made in the literature to date
about limiting $\tan\beta$ using $a_\mu$: at present we are not
able to place any $1\sigma$ bound on $\tan\beta$ larger than 1 due to
the very real possibility of a neutralino-dominated $\delta a_\mu$.
Future reduction in the error bars on $a_\mu$ will provide an
important opportunity for placing a meaningful lower bound on $\tan\beta$.


\subsection{Implications for the Tevatron}

At its simplest level, the measurement of $\delta a_\mu$, an anomaly
in the lepton sector, has little impact on the Tevatron, a hadron
machine. In particular, the light smuons associated with $\delta
a_\mu$ cannot be directly produced at the Tevatron, occuring only if
heavier (non-leptonic) states are produced which then decay to
sleptons. In the calculation of $a_\mu$, the only such sparticles are
the neutralinos and charginos. These states can be copiously produced
and in fact form the initial state for the ``gold-plated'' SUSY
trilepton signature. 

Of particular interest for the trilepton signature are the masses of
the lighter chargino ($\char_1$) and 2nd lightest neutralino ($\neut_2$).
Studies of mSUGRA parameter space indicate that the sensitivity to the
trilepton signature at Run II/III of the Tevatron depends strongly on
the mass of sleptons which can appear in the gaugino decay chains. For
heavy sleptons, the Tevatron is only sensitive to gaugino masses in
the range~\cite{tevsugra}
$m_{\char_1,\neut_2}\lsim 130$ to $140\gev$ for $10\invfb$ of 
luminosity and $145$ to $155\gev$ for $30\invfb$, with smaller values
for smaller $\tan\beta$. However, for light sleptons (below about
$200\gev$) the range is considerably extended, up to gaugino masses
around 190 to $210\gev$. Of course, this latter range is exactly the
one most relevant for understanding $a_\mu$.

\begin{figure}
\centering
\epsfxsize=4.25in
\hspace*{0in}
\epsffile{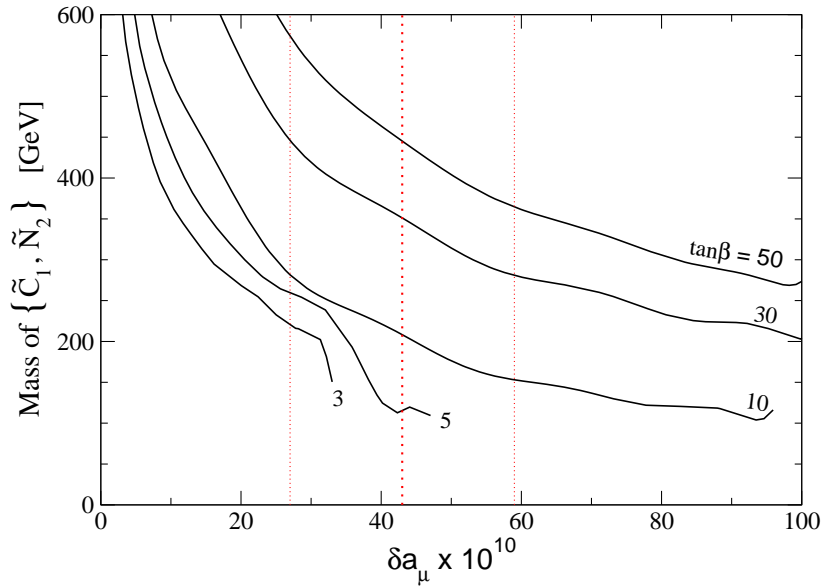}
\caption{Mass bounds on $\char_1$ and $\neut_2$ (where
$m_{\char_1}\simeq m_{\neut_2}$) as a function of
$\delta a_\mu$ for $\tan\beta=3$, 5, 10, 30 and 50. The dotted lines
represent the E821 central value and $1\sigma$ bounds on $\delta a_\mu$.
This figure
assumes gaugino unification and a neutralino LSP. 
}
\label{tevatron}
\end{figure}

It is impossible in the kind of analysis presented here
to comment on the expected
cross-sections for the neutralino-chargino production (there is
no information in $a_\mu$ on the masses of the $t$-channel squarks,
for example)
but we can examine the mass bounds on $\char_1$ and $\neut_2$. In
Fig.~\ref{tevatron} we have shown just that: the upper bound on the
{\em heavier}\/ of either $\char_1$ or $\neut_2$ as a function of
$\delta a_\mu$ for several values of $\tan\beta$. 

A few comments are in order on the plot. First, the plot assumes
gaugino unification; dropping that assumption can lead to
significantly heavier masses for the $\neut_2$ though not for the
$\char_1$; this is because the light $\char_1$ is needed to
participate in the magnetic moment, while $\neut_2$ can
decouple. Second, we have also assumed a neutralino LSP; this is to be
expected since the event topology for the trilepton signal assumes a
stable, neutralino LSP. Finally, on the $y$-axis is actually plotted
$m_{\char_1}$, but in every case we examined with gaugino unification,
the difference in the maximum masses of $\char_1$ and $\neut_2$
differed by at most a few GeV. This is because they are both
dominantly wino-like in the unified case and thus have masses
$\simeq M_2$.

From the figure it is clear that a no-lose theorem for the
Tevatron is not lurking in the current E821 data. However, if the central
value reported by E821 holds up and $\tan\beta\lsim 10$ then one
should expect the Tevatron to find a trilepton signal for SUSY. So
there is actually significant hope for a positive result. We cannot
emphasize enough too that these are {\em upper bounds}\/ on the
sparticle masses and in no way represent best fits or preferred
values. Thus even for larger $\tan\beta$ or smaller $\delta a_\mu$,
there is good reason to hope that the Tevatron will be able to probe
the gaugino sector in Run II or III.

\subsection{Implications for a Linear Collider}

A concensus is currently forming in favor of building a
$\sqrt{s}=500\gev$ linear collider, presumably a factory for
sparticles with masses below $250\gev$. What does the measurement of
$a_\mu$ tell us with regards to our chances for seeing SUSY at
$\sqrt{s}=500\gev$? And how many sparticles will be actually
accessible to such a collider? 

\begin{figure}
\centering
\epsfxsize=4.25in
\hspace*{0in}
\epsffile{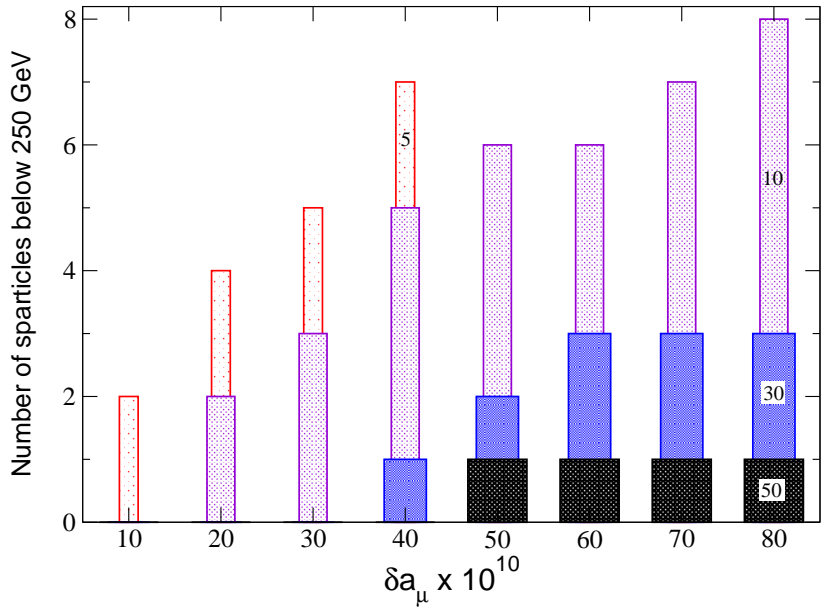}
\caption{Minimum number of sparticles directly observable at a
$\sqrt{s}=500\gev$ linear collider as a function of $\delta a_\mu$. 
The bars represent $\tan\beta=5$, 10, 30 and 50 in order of increasing
thickness. This graph assumes gaugino unification but 
does not include additional sleptons
implied by slepton mass universality.
}
\label{nlchist}
\end{figure}

The analysis of the previous section
can put a lower bound on the number of observable sparticles at a
linear collider as a function of $\delta a_\mu$ and $\tan\beta$ and we
show those numbers as a histogram in Fig.~\ref{nlchist}. In this
figure, we have shown the {\em minimum}\/ number of sparticles with
mass below $250\gev$ for $\tan\beta=5$, 10, 30 and 50, assuming
gaugino unification. In the graph,
the thinner bars represent smaller $\tan\beta$. As is to be
expected, the number of light states increases with increasing
$\delta a_\mu$ and with decreasing $\tan\beta$. However note
that there are no $\tan\beta=5$ lines for $\delta a_\mu>40\times
10^{-10}$ since there is no way to explain such large $\delta a_\mu$
values at low $\tan\beta$.

We see from the figure also that 
for $\tan\beta\gsim30$ there is no guarantee that a
$500\gev$ machine would produce on-shell sparticles; this is not to be
taken to mean that one should not expect their production, simply that
$a_\mu$ cannot guarantee it. However for $\tan\beta\lsim 10$, the
$1\sigma$ limit on $a_\mu$ from E821 indicates that at least 2 to 3
sparticles will be accessible to a $500\gev$ machine. This counting
does not include extra sleptons due to slepton mass universality; for
example, a light muon sneutrino also implies light tau and electron
sneutrinos, and likewise for the charged smuon. 

A similar bar graph can also be made for a $1\tev$ machine, though we
do not show it here. However the relevant numbers can be inferred from
Fig.~\ref{massfig}; we see that such a machine is guaranteed to
produce at least three sparticles for $\tan\beta\leq50$ at the E821
$1\sigma$ limit.

\section{Conclusions}

Deviations in the muon anomalous magnetic moment have long been
advertised as a key hunting ground for indirect signatures of SUSY. And
it may be that we finally have the evidence we need in the
$2.7\sigma$ deviation reported by E821. If this signal is real and if
SUSY is the correct explanation for it, then light sparticles are
requisite. In this analysis, we have confirmed, extended and
overridden some of the bounds on these light sparticles that have
appeared previously. In particular, for the $1\sigma$ lower bound on
$\delta a_\mu$ derived from the E821 data, we obtain the following for
the most general MSSM parametrization:
\begin{itemize}
\item there must be at least 4 sparticles with masses below $820\gev$
($700\gev$ for unified gauginos);
\item the lightest sparticle must lie below $440\gev$ ($345\gev$ for
unified gauginos);
\item there is no lower bound on $\tan\beta$ (the central value
of the E821 data requires $\tan\beta>4.7$);
\item the $\smuon_1$ and the $\sneut$ must lie below $870\gev$ while
one neutralino must fall below $610\gev$ (those bounds become $680$
and $455\gev$ respectively for unified gauginos);
\item there is no upper bound on $m_{\char_1}$ (the data's central
value requires $m_{\char_1}<690\gev$);
\item if $\tan\beta\lsim 10$, then a $500\gev$ linear collider is
guaranteed to produce at least 2 sparticles, and Run~II/III of the
Tevatron is likely to see a trilepton signature (assuming a stable
neutralino LSP); for $\tan\beta\gsim 10$, neither of these can be guaranteed.
\end{itemize}
In the most interesting theoretical scenario, \ie\ gaugino unification
with a neutralino LSP, the bounds on SUSY masses can be much lower,
with the neutralino LSP below $340\gev$, $\char_1$ and $\neut_2$ below
$575\gev$ and sleptons below $825\gev$. Our only important
simplifications were to drop the dependence on $A_\smuon$ from the
smuon mixing matrices, and to assume slepton mass universality.

If E821 has discovered evidence for SUSY, then it is clear that we
have many more discoveries awaiting us. The current data implies that
the LHC will find SUSY directly; such a discovery for the Tevatron or
the NLC is not guaranteed, though the data on $a_\mu$ greatly enhances
the likelihood that these programs will be successful.

\section*{Acknowledgments}

We are grateful to S.~Martin for discussions during the early parts of
this project. CK would also like to thank the Aspen Center for Physics
where parts of this work were completed, and the Notre Dame
High-Performance Computing Cluster for much-needed computing
resources. This work was supported in part by the National Science
Foundation under grant NSF-0098791. The work of MB was supported in
part by a Notre Dame Center for Applied Mathematics Graduate Summer
Fellowship.

\section*{Appendix}

The supersymmetric contributions to $a_\mu$ are generated by diagrams
involving charged smuons with neutralinos, and sneutrinos with
charginos. The most general form of the calculation, including phases,
takes the form (we follow Ref.~\cite{martinwells}):
\bea
\delta a_\mu^{(\neut)} & = & \frac{m_\mu}{16\pi^2}
   \sum_{i,m}\left\{ -\frac{m_\mu}{ 12 m^2_{\smuon_m}}
  (|n^L_{im}|^2+ |n^R_{im}|^2)F^N_1(x_{im}) 
 +\frac{m_{\neut_i}}{3 m^2_{\smuon_m}}
    {\rm Re}[n^L_{im}n^R_{im}] F^N_2(x_{im})\right\}
\nonumber \\
\delta a_{\mu}^{(\char)} & = & \frac{m_\mu}{16\pi^2}\sum_k
  \left\{ \frac{m_\mu}{ 12 m^2_{\sneut}}
   (|c^L_k|^2+ |c^R_k|^2)F^C_1(x_k)
 +\frac{2m_{\char_k}}{3m^2_{\sneut}}
         {\rm Re}[ c^L_kc^R_k] F^C_2(x_k)\right\}
\eea
where $i=1,2,3,4$, $m=1,2$, and $k=1,2$ label the neutralino, smuon
and chargino mass eigenstates respectively, and
\bea
n^R_{im} & = &  \sqrt{2} g_1 N_{i1} X_{m2} + y_\mu N_{i3} X_{m1},
\nonumber \\
n^L_{im} & = &  {1\over \sqrt{2}} \left (g_2 N_{i2} + g_1 N_{i1}
\right ) X_{m1}^* - y_\mu N_{i3} X^*_{m2}, \nonumber \\
c^R_k & = & y_\mu U_{k2}, \nonumber \\
c^L_k & = & -g_2V_{k1},
\eea
$y_\mu = g_2 m_\mu/\sqrt{2} m_W \cos\beta$ is the muon Yukawa
coupling, and $g_{1,2}$ are the U(1) hypercharge and SU(2) gauge couplings.
The loop functions depend on the
variables $x_{im}=m^2_{\neut_i}/m^2_{\smuon_m}$, 
$x_k=m^2_{\char_k}/m^2_{\sneut}$ and are given by
\bea
F^N_1(x) & = &\frac{2}{(1-x)^4}\left[ 1-6x+3x^2+2x^3-6x^2\ln x\right] 
\nonumber \\
F^N_2(x) & = &\frac{3}{(1-x)^3}\left[ 1-x^2+2x\ln x\right] 
\nonumber \\
F^C_1(x) & = &\frac{2}{(1-x)^4}\left[ 2+ 3x - 6
x^2 + x^3 +6x\ln x\right] \nonumber \\
F^C_2(x) & = &-\frac{3}{2(1-x)^3}\left[ 3-4x+x^2 +2\ln x\right]
\eea
For degenerate sparticles ($x=1$) the functions are normalized so that
$F^N_1(1) = F^N_2(1) = F^C_1(1) = F^C_2(1) = 1$. We can also bound the
magnitude of some of these functions; in particular $|F^N_2(x)|\leq 3$
while $|F^C_2(x)|$ is unbounded as $x\to0$.

The neutralino and chargino mass matrices are given by
\beq
M_{\neut} = \left(\begin{array}{cccc} M_1 & 0 & - m_Z s_W c_\beta &
m_Z s_W s_\beta \\
0 & M_2 & m_Z c_W c_\beta & - m_Z c_W s_\beta \\
-m_Z s_W c_\beta & m_Z c_W c_\beta & 0 & -\mu \\
m_Z s_W s_\beta & - m_Z c_W s_\beta& -\mu & 0 \end{array} \right)
\label{neutralinomassmatrix} 
\eeq
and
\beq
M_{\char} = 
\left(\begin{array}{cc} M_2 & \sqrt{2} m_W s_\beta \\
                              \sqrt{2} m_W c_\beta & \mu \end{array}
	\right)
\label{charginomassmatrix}
\eeq
where $s_\beta = \sin\beta$,
$c_\beta = \cos\beta$ and likewise for $\theta_W$.
The
neutralino
mixing matrix 
$N_{ij}$ 
and the chargino mixing  matrices $U_{kl}$ and $V_{kl}$ satisfy
\bea
N^* M_{\neut} N^\dagger &=& {\rm diag}(
m_{\neut_1},
m_{\neut_2},
m_{\neut_3},
m_{\neut_4}) \nonumber\\
U^* {M}_{\char} V^\dagger &=& {\rm diag}(
m_{\char_1},
m_{\char_2})
. 
\eea

The smuon mass matrix is given in the 
$\{ \smuon_L, \smuon_R \}$ basis as:
\beq
M^2_{\smuon}=\left( \begin{array}{cc}
	M_L^2 & m_\mu (A^*_{\tilde\mu}-\mu\tan\beta) \\
	m_\mu (A_{\tilde\mu}-\mu^*\tan\beta) & M_R^2 \end{array}
	\right)
\eeq
where
\bea
M_L^2 & = & m^2_L +(s_W^2 -\frac{1}{2})m_Z^2\cos 2\beta \nonumber
\\
M_R^2 & = &  m^2_R -s_W^2 \, m_Z^2\cos 2\beta 
\eea
for soft masses $m_L^2$ and $m_R^2$;
the unitary smuon mixing matrix $X_{mn}$ is defined by
\bea
X M^2_{\smuon}\, X^\dagger = 
{\rm diag}\, (m^2_{\smuon_1}, m^2_{\smuon_2}).
\eea
We will define a smuon mixing angle $\theta_\smuon$ such that 
$X_{11}=\cos\theta_\smuon$ and $X_{12}=\sin\theta_\smuon$. In our
numerical calculations we will set $A_\smuon=0$. At low $\tan\beta$ we
have checked that varying $A_\smuon$ 
makes only slight numerical difference, while at large
$\tan\beta$ it has no observable effect whatsoever. Finally, 
the muon sneutrino mass is related to the left-handed smuon mass parameter
by
\beq
m_{\sneut}^2 = m_L^2 + {1\over 2} m_Z^2 \cos 2\beta  .
\eeq

The leading 2-loop contributions to $\delta a_\mu$ have been
calculated~\cite{2loop} and have been found to suppress the SUSY
contribution by a factor $(4\alpha/\pi)\log(m_{\rm
SUSY}/m_\mu)\approx 0.07$; we will include this 7\% suppression in all
our numerical results.


\begin{thebibliography}{99}

\bibitem{e821}
H.~N.~Brown {\it et al.}  [Muon g-2 Collaboration],
Phys.\ Rev.\ Lett.\  {\bf 86}, 2227 (2001).

\bibitem{czar}
A.~Czarnecki and W.~J.~Marciano,
Phys.\ Rev.\ D {\bf 64}, 013014 (2001).

\bibitem{kane}
L.~Everett, G.~Kane, S.~Rigolin and L.~Wang,
Phys.\ Rev.\ Lett.\  {\bf 86}, 3484 (2001).

\bibitem{feng}
J.~L.~Feng and K.~T.~Matchev,
Phys.\ Rev.\ Lett.\  {\bf 86}, 3480 (2001).

\bibitem{martinwells}
S.~P.~Martin and J.~D.~Wells,
Phys.\ Rev.\ D {\bf 64}, 035003 (2001).

\bibitem{others}
U.~Chattopadhyay and P.~Nath,
Phys.\ Rev.\ Lett.\  {\bf 86}, 5854 (2001); \\
S.~Komine, T.~Moroi and M.~Yamaguchi,
Phys.\ Lett.\ B {\bf 506}, 93 (2001) and
Phys.\ Lett.\ B {\bf 507}, 224 (2001); \\
J.~R.~Ellis, D.~V.~Nanopoulos and K.~A.~Olive,
Phys.\ Lett.\ B {\bf 508}, 65 (2001);\\
R.~Arnowitt, B.~Dutta, B.~Hu and Y.~Santoso,
Phys.\ Lett.\ B {\bf 505}, 177 (2001); \\
K.~Choi \etal,
Phys.\ Rev.\ D {\bf 64}, 055001 (2001); \\
J.~E.~Kim, B.~Kyae and H.~M.~Lee,
hep-ph/0103054; \\
K.~Cheung, C.~Chou and O.~C.~Kong,
hep-ph/0103183; \\
H.~Baer \etal,
Phys.\ Rev.\ D {\bf 64}, 035004 (2001); \\
F.~Richard,
hep-ph/0104106; \\
C.~Chen and C.~Q.~Geng,
Phys.\ Lett.\ B {\bf 511}, 77 (2001); \\
K.~Enqvist, E.~Gabrielli and K.~Huitu,
Phys.\ Lett.\ B {\bf 512}, 107 (2001);\\
D.~G.~Cerdeno \etal, 
hep-ph/0104242; \\
G.~Cho and K.~Hagiwara,
Phys.\ Lett.\ B {\bf 514}, 123 (2001).

\bibitem{davier}
M.~Davier and A.~H\"ocker,
Phys.\ Lett.\ B {\bf 435}, 427 (1998).

\bibitem{yndurain}
F.~J.~Yndurain,
hep-ph/0102312.

\bibitem{marciano}
W.~J.~Marciano and B.~Lee~Roberts,
hep-ph/0105056.

\bibitem{calcs}
J.~A.~Grifols and A.~Mendez,
Phys.\ Rev.\ D {\bf 26}, 1809 (1982);\\
R.~Barbieri and L.~Maiani,
Phys.\ Lett.\ B {\bf 117}, 203 (1982);\\
D.~A.~Kosower, L.~M.~Krauss and N.~Sakai,
Phys.\ Lett.\ B {\bf 133}, 305 (1983);\\
T.~C.~Yuan \etal, 
Z.\ Phys.\ C {\bf 26}, 407 (1984);\\
U.~Chattopadhyay and P.~Nath,
Phys.\ Rev.\ D {\bf 53}, 1648 (1996); \\
T.~Moroi,
Phys.\ Rev.\ D {\bf 53}, 6565 (1996),
[{\bf E}:~{\it ibid.}\ D {\bf 56}, 4424 (1996)];\\
M.~Carena, G.~F.~Giudice and C.~E.~Wagner,
Phys.\ Lett.\ B {\bf 390}, 234 (1997);\\
T.~Ibrahim and P.~Nath,
Phys.\ Rev.\ D {\bf 61}, 095008 (2000).

\bibitem{haberkane}
H.~E.~Haber and G.~L.~Kane,
Phys.\ Rept.\  {\bf 117}, 75 (1985).

\bibitem{tevsugra}
For a review, see:\\
S.~Abel {\it et al.}  [Tevatron Run~II SUGRA Working Group],
hep-ph/0003154.

\bibitem{2loop}
G.~Degrassi and G.~F.~Giudice,
Phys.\ Rev.\ D {\bf 58}, 053007 (1998).


\end{thebibliography}
\end{document}